# OY Car in Outburst:

## Balmer Emission From The Red Star And The Gas Stream

E. T. Harlaftis[1] and T. R. Marsh[2]

[1] Isaac Newton Group of Telescopes, Apartado de Correos 321,
Santa Cruz de La Palma, E-38780 Canary Islands, Spain.
ehh@st-and.ac.uk
[2] Department of Physics, University of Southampton,
Southampton, SO17 1BJ, UK.
trm@phastr.soton.ac.uk

**Abstract.** We present observations of OY Car, obtained with the Anglo-Australian Telescope, during a normal outburst in August 1991. Two sinusoidal components are resolved in the H$\beta$ trailed spectra and we determine the location of the narrow component to be on the secondary star with a maximum contributed flux of $\sim$2.5 per cent to the total flux. Imaging of the line distribution reveals that the other emission component is associated with the gas stream. This follows a velocity close to the ballistic one from the red star to a distance of $\sim$0.5 $R_{L_1}$ from the white dwarf. Then, its kinematics changes from 0.5–0.2 $R_{L_1}$ (accretion disc) following velocities now closer to (but lower than) the keplerian velocities along the path of the gas stream. We finally discuss the implications of having observed simultaneously line emission from the gas stream and the red star during outburst.

**Key words:** stars: binaries: eclipsing – stars: cataclysmic variables – stars: individual: OY Car

## 1. Introduction

The secondary stars are often undetectable in cataclysmic variables. This is particularly true for systems of short orbital period - in which the secondary stars are especially dim (M dwarfs) - and for high mass transfer systems where light from the accretion disc dominates (nova-like variables). In recent years, line emission originating from the red star has been detected in a few cataclysmic variables (UX UMa, Schlegel et al. 1983; RW Tri, Kaitchuck et al. 1983; SS Cyg, Hessman 1986; IX Vel, Beuermann & Thomas 1990; RX And, Kaitchuck, Mansperger and Hantzios 1988; IP Peg, Hessman 1989; IP Peg, Marsh and Horne 1990; U Gem, Marsh et al. 1990; IP Peg, Harlaftis et al. 1994; DW UMa, Dhillon et al. 1994; YZ Cnc, Harlaftis et al. 1995) providing a new approach in studying the secondary star. Indeed, there are cases where the secondary star can be detected and subsequently studied only by its line emission (e.g. IX Vel; Beuermann and Thomas 1990). We realized that such narrow emission lines from the secondary star may not have been resolved before and that, with adequate resolution, these lines could be detected in the trailed spectra of many more systems than previously thought. Here, we present the first detection of an emission component from the secondary star of OY Car which was made possible from the higher than normal resolution of our observations and the outburst state of the system.

## 2. Observations and data reduction

OY Car was observed with the 3.9-m Anglo-Australian Telescope for a full orbital cycle on the night 31 August 1991 while the target was at the maximum of a normal outburst (maximum was 12.1 mag on 30 August and lasted for one day; Frank Bateson, private communication). The orbital cycle was covered with 78 spectra of 1 minute long exposure (dead time between exposures was 10 seconds). The RGO spectrograph with the Thomson CCD and a grating of 1200 lines mm$^{-1}$ gave a wavelength coverage from 4585–4965 Å at FWHM of 1.26 Å (or $\sim$80 km s$^{-1}$ at H$\beta$) resolution. The slit was rotated so that a comparison star was placed in the slit for later slit loss corrections (seeing was 2.3 arcseconds). Wide slit photometric spec-

*Send offprint requests to*: E. T. Harlaftis
\* *Present address:* University of St. Andrews, School of Physics and Astronomy, North Haugh, St Andrews, Fife KY16 9SS, U.K.

also obtained. Comparison arc spectra were taken every 30 min. After the bias level was removed from the data images and the sky contribution subtracted from the spectra, the object and comparison spectra were optimally extracted (Marsh 1989). For the wavelength calibration of the spectra, interpolation was used between neighbouring arc spectra. The root mean square of the polynomial fits is $\sim 0.014$ Å. Although we corrected for slit losses, the flux calibration is only accurate to 25 per cent mainly because some observations were seriously affected by clouds and high air-mass (the latter resulted because the observations were done in an unfavourable month for OY Car). The ephemeris used for the orbital phase determination is

HJD = 2443993.552733 + 0.0631209239 E

as derived from Space Telescope observations of OY Car and is accurate to $\sim 1$ second (Marsh et al. 1995, in preparation).

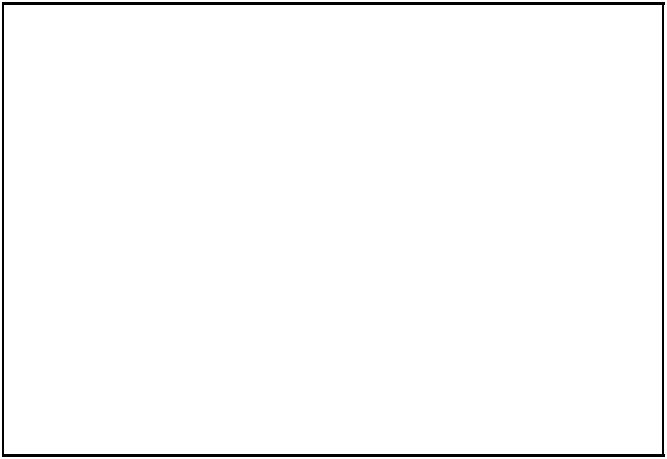

**Fig. 1.** The average spectrum of OY Car out-of-eclipse showing the complex emission lines during outburst

### 2.1. The Average Spectrum

We display the average out-of-eclipse spectrum of OY Car in Fig. 1. The blue spectrum shows double-peaked H$\beta$ and He I emission profiles with prominent absorption cores going below the continuum level. High excitation lines such as the C III/N III and He II $\lambda$4686 are in emission, apparently forming double-peaked profiles with an absorption core component. This looks similar to spectra of IP Peg during outburst although there He II 4686 is stronger than the H$\beta$ (Marsh & Horne 1990). This is possibly related to the higher mass-transfer rate observed in long period cataclysmic variables which increases the temperature of the photoionizing source. Unfortunately, there are no other optical spectra of OY Car or Z Cha in outburst and very few in superoutburst to compare with.

show a stronger red peak to the blue one (which was attributed by the authors to a precessing disc) and an absorption core which is well above the continuum level (Hessman et al. 1992). Spectra of Z Cha in superoutburst show the same characteristics as OY Car in outburst (Vogt 1982). A comparison with spectra of OY Car during quiescence shows that the main difference is the absence of the white dwarf's broad absorption line wings in the Balmer profiles (see Fig. 1 in Hessman et al. 1989) and the stronger absorption cores of the He I lines at $\lambda$4713 and $\lambda$4922 during outburst. These differences can be attributed to an increased optical thickness of the disc during outburst which would render the photospheric lines from the white dwarf undetectable (i.e. the broad absorption wings) while it would increase the strength of the observed absorption core.

### 2.2. Light Curves

We rebinned the spectra onto a uniform wavelength scale and averaged them into 10 binary phase bins. We subtracted each individual continuum from each spectrum using a fitting procedure, resulting in the graph displayed in Fig. 2. There is orbital variation in the line profiles, most notably in H$\beta$. The blue peak of the profile is clearly stronger than the red peak between phases 0.4-0.7. For these phases, the minimum of the absorption core has a redward offset of $\sim 100$ km s$^{-1}$ relative to the centre of the line. In addition, we observe that the emission peak closer to zero velocity is the stronger one (blue or red, depending on phase).

During eclipse (−0.03 to +0.03 orbital phase; in quiescence both white dwarf and bright spot are eclipsed; Wood et al. 1989a), the width and height of the profile peaks are identical within the errors (using Gaussian profile fits). However, the Gaussian centres of the peaks are −510±50 and +860±50 km s$^{-1}$ which indicate a redward $\gamma$ velocity shift for the eclipse profile. Such velocity shifts have been linked to sinusoidal components present in the trailed spectra (Hessman et al. 1989; see also next section of this paper).

The eclipse spectrum shows clear evidence of a double-peaked profile for He I $\lambda$4922 and possibly for He II. The H$\beta$ eclipse profile shows a symmetrical double-peaked profile with the absorption core above the continuum. Since the inner disc is eclipsed, the velocity separation of the two peaks can give an estimate of the outer disc radius, assuming keplerian velocities, from the equation

$$\frac{R_{disc}}{a} = (1+q) \left(\frac{K_R}{V_{min}}\right)^2$$

where $2 V_{min} = 1390\pm30$ km s$^{-1}$ from Fig. 2, (typical of high inclination systems; Bailey & Ward 1981). For the above calculation we used results from Wood & Horne (1990): $K_R = 460$ km s$^{-1}$; a limb-darkening parameter of $0 < u < 1$; and $q = 0.102$. This separation gives $R_{disc}$

**Fig. 2.** The spectra of OY Car in outburst are shown with orbital phase (after their continuum has been subtracted; this simplifies comparison with the eclipse spectrum). The data have been averaged into 10 bins and have been offset in the y-direction for clarity by 6 mJy each. The dominant lines are marked. Note the emission flux in H$\beta$ is maximum and the absorption component is minimum during eclipse

= 0.48±0.02 $a$, which is larger than the bright spot radius during quiescence, $R_{bs}$ = 0.31±0.02 $a$ (where $a$ is the binary separation; Wood et al. 1989a). This is consistent with a disc radius increase during outburst (Wood et al. 1989b) although sub-keplerian velocities in the outer disc may also contribute to a difference between the bright spot radius and the emission line region (Wade & Horne 1988). Finally, the equivalent width of H$\beta$ is around 7-8 Å and drops slightly to 5-6 Å at orbital phase 0.9. During eclipse it rises up to 75Å. Thus, the effect of the orbital hump is not very evident in the light curve of the H$\beta$ equivalent width and can account at most for 25 per cent. Although the emission flux calibration is not very accurate, no significant emission line flux in H$\beta$ is evident from the orbital hump.

### 2.3. The Trailed Spectra

The trailed spectra are shown as an image in Fig. 3 (78 spectra in total) after having subtracted the continuum with a spline function. Close inspection of the double-peaked spectra reveals various features caused by (a) the eclipse, (b) a sinusoidal wave from the red star and (c) an 'S'-wave. During eclipse, the classical rotational disturbance or 'z'-wave is seen (blue peak is first eclipsed and then emerges first out of the eclipse), evidence of a prograde accretion disc been eclipsed (Greenstein & Kraft 1959). During eclipse, the absorption component disap- (defined by the absorption core reaching the continuum level) at phases 0.957-0.984 and 1.012-1.039. We can constrain the size of the absorption region from the phases that it is observed to be fully eclipsed. The phases over which the white dwarf is fully eclipsed are from 0.980 to 1.020 (see Fig. 3 from Wood et al. 1989a) which are consistent with the range over which the absorption component disappears (i.e. comparable size), and thus its origin is related to the material which lies in the line of sight to the white dwarf. A sharp, sinusoidal, emission component which moves in the opposite direction to that of the double-peaked profiles (i.e. the accretion disc) is also evident. In particular, it crosses the Balmer peaks (red to blue) at around phase 0.5, suggesting that its origin is related to the secondary star. More evidence on this is given in the following section.

It is possible to trace part of another sinusoidal-like, emission component in Fig. 3 with a larger velocity amplitude than the emission component we have already discussed. It appears strong on the red peak after the eclipse (0.1-0.2 phase or ∼36 degrees), then disappears and reemerges strongly in the blue peak at phases 0.5-0.6, crosses the profile at phase ∼0.75 and thereafter can be traced back to eclipse. It is apparent that these characteristics are signs of an 'S'-wave trailing the motion of the red star by about 0.25 cycles More evidence of the presence of the 'S'-wave is shown in the next section. The trailed spectra of HeI $\lambda$4922 show weak disc emission with less extended wings than H$\beta$ and HeII. The structure of the absorption core with phase is similar to H$\beta$ (e.g. is eclipsed) and finally there is a hint of the 'S'-wave at around phase 0.1.

### 3. Doppler Tomography

We used the trailed spectra of the H$\beta$ line to apply the indirect imaging technique of Doppler tomography (Marsh and Horne 1988). This is a maximum entropy method (MEM) of reconstructing the emission-line distribution in a cataclysmic variable from the orbital line-profile variations. In a Doppler image, the intensity of a pixel at a velocity ($V_X$, $V_Y$) corresponds to the intensity of the appropriate sinusoidal component in the trailed spectra

$$V = \gamma - V_X \cos(2\pi\phi) + V_Y \sin(2\pi\phi)$$

The semiamplitude of each sinusoidal component is given by the distance from the mass centre, while its phase is estimated from the azimuthal angle from the line-of-centres in the map. For example, emission from the red star follows $V_R = K_R \sin(2\pi\phi)$, and appears on the image at (0,$K_R$). Examples of the application to real data are given in Marsh & Horne (1990), Marsh et al. (1990), Dhillon et al. (1994) and Harlaftis et al. (1994).

It was possible to apply the technique only to the H$\beta$ line because of signal-to-noise limitations. However, the H$\beta$ profiles after subtraction of the continuum have many negative values because of the absorption core but the im-fore, we added a Gaussian of FWHM = 700 km s$^{-1}$ to the data so that all have positive values, and then after the reconstruction of the image we subtracted the equivalent gaussian image in order to produce the final result. This procedure has no effect on the goodness of the fit or entropy when the latter is determined over small scales (Marsh et al. 1990). Image structure is gradually built on a default image (starting from a uniform image) by reducing the $\chi^2$ between the data and the model fit. When $\chi^2$ is reaching its optimum value (a level before noise becomes apparent in the image), then the solution is found by maximizing the entropy in the image. We used a Gaussian blurring of FWHM=10 pixels on the default image (see Appendix). We further minimized $\chi^2$ (=1.35) by adjusting the systemic velocity which can blur the image (Marsh & Horne 1988). We obtained $\gamma$=+29 km s$^{-1}$.

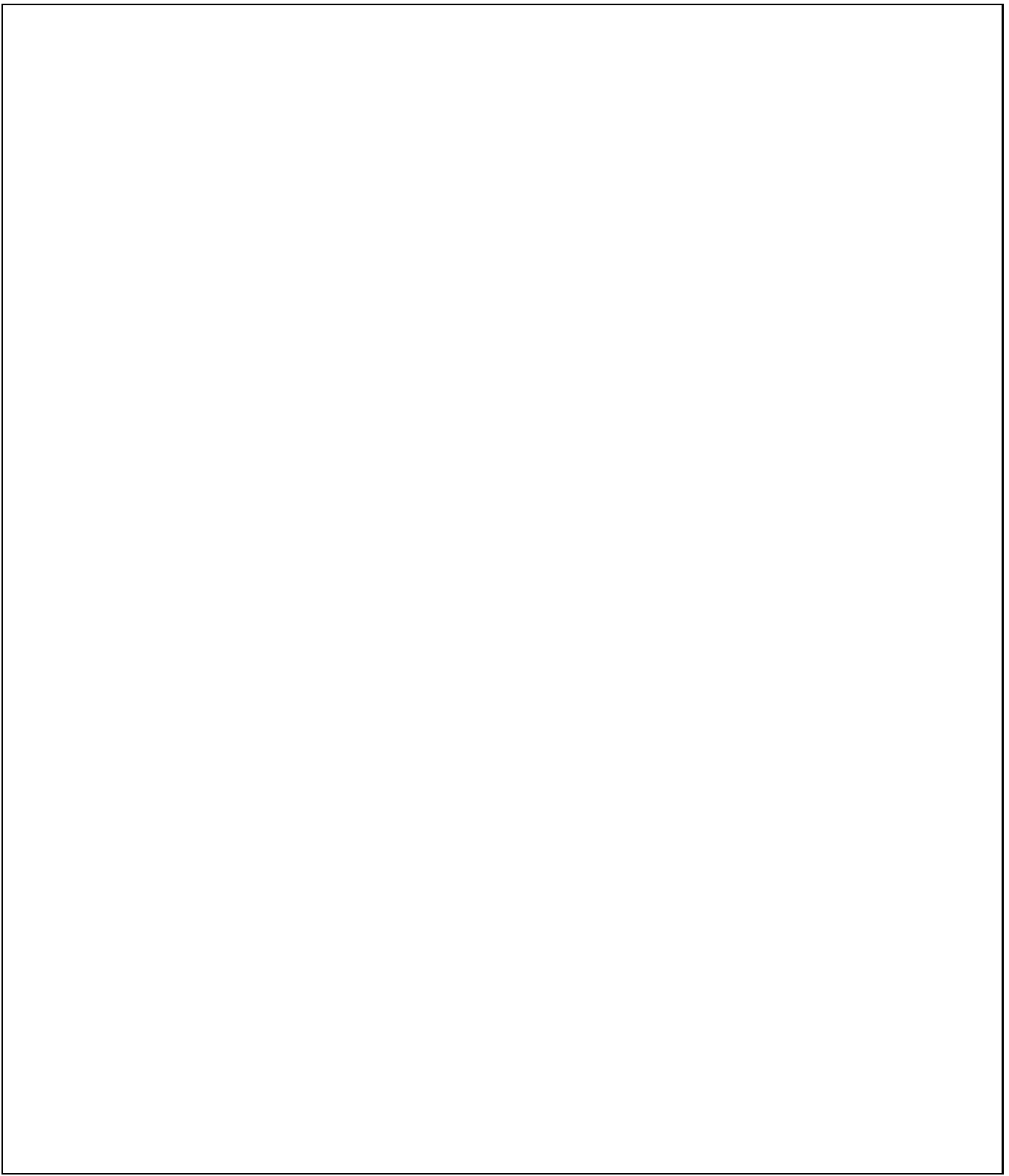

**Fig. 3.** The trailed spectra of H$\beta$ in OY Car (78 AAT spectra). The line He I 4921 is also displayed. The continuum has been subtracted from the spectra. Note the eclipse, the narrow sinusoidal component crossing red to blue at around phase 0.5 and parts of an 'S'-wave. See the text for more details

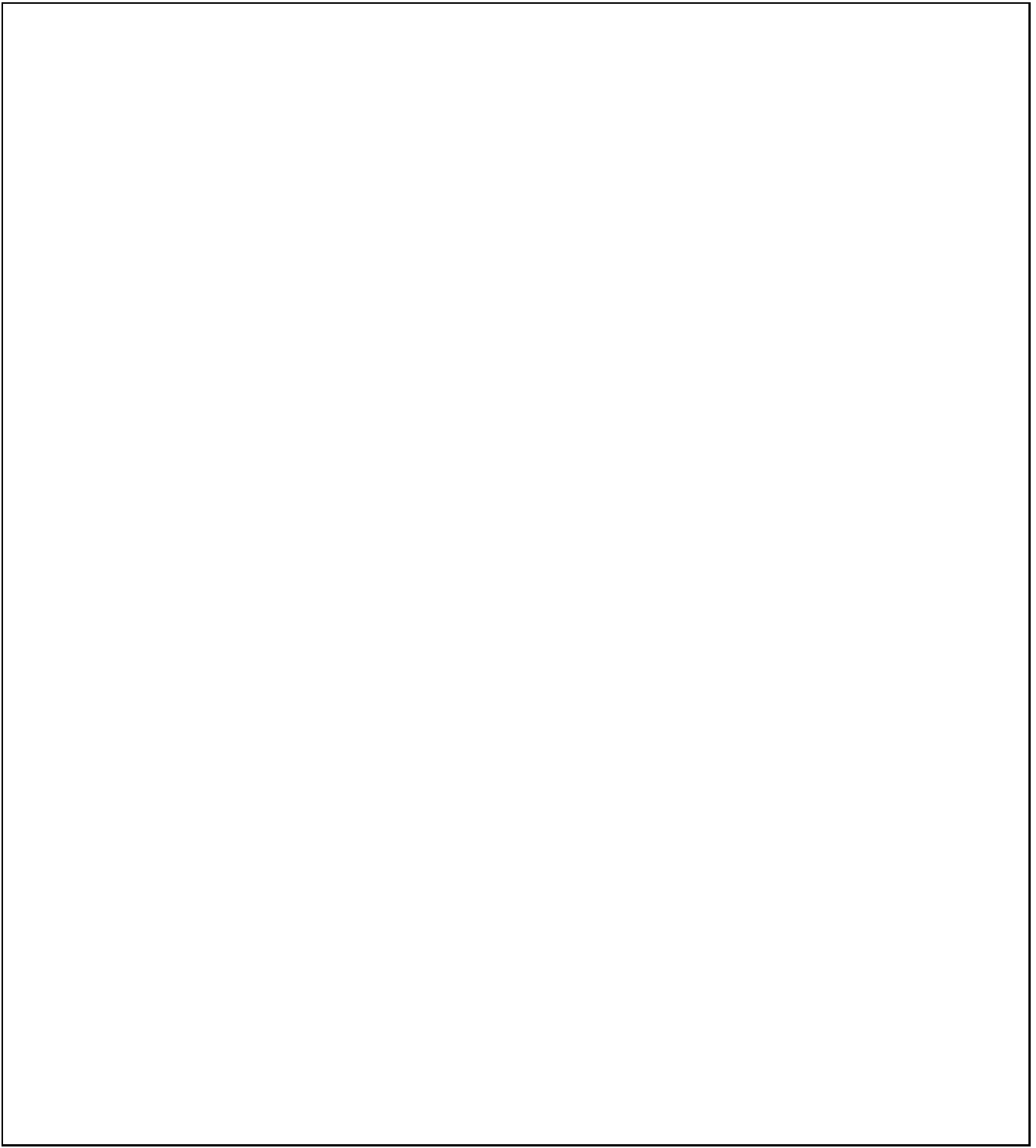

**Fig. 4.** The Doppler tomography analysis. The trailed spectra of H$\beta$ in OY Car are displayed in panel "a" (top). Panel "b" (top) shows the Doppler map fitted to these data. Panel "c" (top) shows the fits computed from the Doppler map (predicted data). Panel "b" (bottom) displays the non-axisymmetric component of emission. The bottom panels "a" and "c" show the observed and computed non-axisymmetric spectra, respectively. All panels are in the same grey scale. See text for more details.

Fig. 4 shows the results of this analysis for OY Car (common intensity scale of 0.02–0.8 mJy). The trailed spectra (54 out-of-eclipse spectra in the orbital range 0.1–0.9) of H$\beta$ are displayed in top panel "a" (with the velocity relative to the line centre along the horizontal axis and orbital phase along the vertical axis). Top panel "b" shows the Doppler map fitted to the observed data. The double-peak profiles, characteristic of emission-line formation in an accretion disc, are evident in the trailed spectra. Reconstruction of the emission-line distribution shows this as a ring; the inner side of the ring (low velocities) corresponds to emission from the outer disc and the outer side of the ring (high velocities) to the region of the accretion disc near the compact object. The image intensity has a velocity-dependence of $V^{-3.4\pm0.2}$ between 800-1400 km s$^{-1}$ which corresponds to $R^{-1.3}$ for a keplerian field. The surface brightness of H$\beta$ with radius is typical of cataclysmic variable stars. Other than the dominant accretion disc emission, a narrow emission component is also evident in the trailed spectra. It crosses the double-peaked profile at phase 0.5 (from the red peak) in anti-phase with the orbital motion. The Doppler map determines the location of this emission-line component to be on the red star.

From the Doppler map, we subtract the axisymmetric Doppler map centred on the white dwarf (0, $-q\ K_R$), formed by the median value at each radius. The result is presented in bottom panel "b" which shows the asymmetries in the line distribution more clearly. Top panel "c" shows the fits computed from the Doppler map (predicted data) and bottom panel "c" shows the fits computed from the non-axisymmetric Doppler map in the same grey scale as the observed data. The computed data (panel "c", overall and asymmetric parts) show that the observed data can be reconstructed reasonably well, including the red star component and the parts of the 'S'-wave where emission is strong (the red peak at phases 0.1-0.3 and blue peak at phases 0.5-0.8). Finally, bottom panel "a" shows the result after the data corresponding to the axisymmetric image has been subtracted from the observed spectra to the phases 0.62–0.67 when the flux standard star was observed (the eclipse was left out of the fit because it represents a geometric effect and the velocity map is intended to be position-independent).

## 4. The non-axisymmetric part of the Doppler image

We re-display the non-axisymmetric part of the Doppler image in Fig. 5 for clarity (higher image definition than in Fig. 4). From top to bottom, the predicted positions of the red star, the centre of mass and the white dwarf are marked as crosses. The positions were determined for $K_R$=460 km s$^{-1}$ (0< $u$ <1), and $q$ = 0.102 (Wood & Horne 1990). The two curves starting from the red star represent the velocity of the gas stream (lower curve) and the keplerian velocity of the disc along the path of the gas stream. These curves are plotted for half orbit of the stream. The open circles denote distance from the white dwarf in units of 0.1 $R_{L_1}$. The asterisk marks the turning point of the velocities from high to low (closest to white dwarf) at a distance of only 0.1 $R_{L_1}$ from the white dwarf. The non-axisymmetric image of H$\beta$ is dominated by extended, arm-like emission which starts from the red star and ends close to the white dwarf, suggesting a deep penetration of the accretion disc by the gas stream.

Some of the emission spreads outside the Roche lobe of the red star and the question arises if this emission comes from the red star. Thereafter, we performed a simulation in order to clarify the effect of the procedure we followed on the construction of the image and distinguish real structures from artifacts (see Appendix). We find that MEM is causing the above effect by blurring (or smoothing) the gas stream emission. However, this blurring does not affect our discussion below since the kinematic or flux information along the path is preserved. In addition, it is likely that there is intrinsic broadening contributing to the gas stream width which is indicated from the trailed spectra. Fig. 3 shows that the 'S'-waves of the red star and the gas stream are not resolved into separate components between phases 0.1–0.2, as is the case with the model trailed spectra in Fig. 6 suggesting that there is a broadening contribution to the gas stream emission larger than the velocity resolution of 80 km s$^{-1}$.

The main arm-like emission between the red star (1.0 $R_{L_1}$) and the 0.5 $R_{L_1}$ point lies close to the velocities of the gas stream or more accurately between the velocities of the gas stream and the disc tracking the stream. The arm-like emission after the 0.5 $R_{L_1}$ point does not follow the gas stream velocity. It follows a path similar to the path of the disc velocities along the trajectory of the stream as far as 0.2 $R_{L_1}$ but with velocities lower by ∼200 km s$^{-1}$. There is no clear evidence for a bright spot, except perhaps the indication given by the above change of the stream velocity appearing at around 0.5 $R_{L_1}$. Note

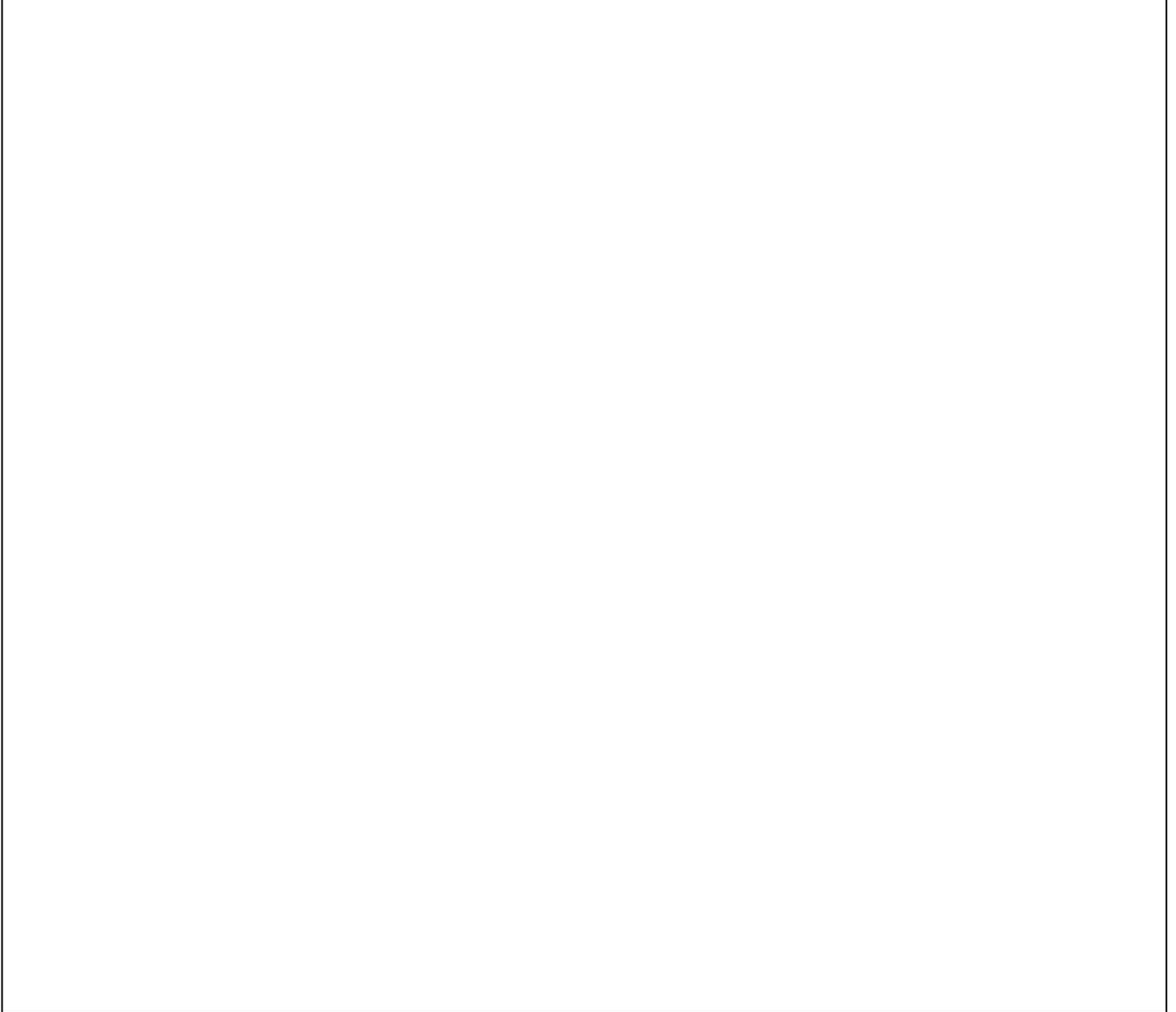

**Fig. 5.** The non-axisymmetric part of the Doppler image expanded for clarity. The red star emission can be resolved coming from its inner face. Theoretical trajectories related to the gas stream are also plotted. See text for more details

that a disc radius of 0.5 $R_{L_1}$ is consistent with the size of the emission line region estimated from the double-peak separation in eclipse. Finally, the peak of the line distribution in Fig. 5 lies in the inner face of the secondary star. A spot-like emission from the red star is not significantly blurred by MEM. Finally, the reliability of the above emission structures is confirmed from the filtered back-projection image (see Appendix).

We estimate the contribution of the secondary star to the H$\beta$ line emission from the non-axisymmetric map to be between 2.4 per cent (counting only pixels covering the Roche lobe of red star) to a minimum of 0.3 per cent (counting only pixels covering the inner face of the red star; assuming that there is underlying emission from the gas stream we obtain a lower limit by subtracting it). The total non-axisymmetric emission accounts for ∼9 per cent of the total H$\beta$ emission which indicates that the gas stream emission is between 7–9 per cent. A direct consequence of the detection of the secondary star from its Balmer emission during outburst is an estimate of its radial velocity which can thus provide an alternative estimate of the $K_R$ binary parameter. In this case where the geometry of the system is well determined we can check the consistency of the above proposed method. The line emission distribution on the red star (after subtracting the underlying blurred emission) is a 3-pixel box between the y-axis) and has its peak and average at the middle pixel centred at $K_V \sim 380$ km s$^{-1}$. Thus, the radial velocity $K_V$ corresponding to this light-center is an under-estimate of the true $K_R$ of the red star ($455 < K_R < 470$ km s$^{-1}$; Wood & Horne 1990). Using a simple irradiation model for the secondary star of OY Car, we find that $\frac{K_V}{K_R} = 0.8$ which is consistent with the above values of radial velocities. (For the mass ratio, binary separation, inclination and distance of OY Car - see Wood & Horne 1990 - we defined a grid of elements over the face of the red star by specifying 65 angle increments for the grid and assumed an irradiating, isotropic source at the white dwarf. We assumed a mean temperature for the red star of 2500 K, a gravity darkening coefficient of 0.08, a linear limb darkening of 0.5, a disk radius of 0.5 $R_{L_1}$ and a vertical extent of only H/R=0.02. For more details of this model see Marsh 1988). It is clear that the limit of this spectroscopic method is the instrumental resolution (80 km s$^{-1}$ in our case).

## 5. Discussion

Emission from the H$\beta$ line is shown to be concentrated towards the inner hemisphere of the red dwarf in OY Car whereas this is not the case for the red star in IP Peg where the line emission is almost symmetrically distributed around the pole (Harlaftis et al. 1994, Marsh & Horne 1990). Geometrical considerations such as the larger size of the red star and the larger (i.e. thicker) disc in IP Peg in comparison to OY Car could in principle explain the above difference. Balmer emission from the secondary star has been observed from about ten dwarf novae in outburst and nova-like variables. Recently, Balmer emission from the red star was detected during outburst in YZ Cnc as well. A Doppler tomography analysis of observations that covered 3 outbursts of YZ Cnc (Harlaftis et al. 1995) showed that a small portion of the H$\alpha$ emission originates on the secondary star, lasting throughout the outburst but dropping with the continuum decline. Marsh and Horne (1990) found in IP Peg Balmer emission from the gas stream during quiescence, and Balmer emission from the secondary star during outburst, in addition to the dominant disc emission. They argued that during outburst this emission originates from the chromosphere of the secondary which is irradiated by the boundary layer between the white dwarf and the inner accretion disc. Soft X-ray and EUV radiation from the boundary layer that is not absorbed by the accretion disc will reach the red star. This effectively means that only the polar regions of the red star will be irradiated, the extent of which will depend on the disc's thickness which can thus be investigated. The weak He II 4686 emission in OY Car shows a double-peak profile (Fig. 2) which suggests that there is a radiation field sufficiently strong to irradiate the disc and possibly the red star. The line emission from the bright spot if not absent (see non-axisymmetric map) is very weak to ac-or for a clear manifestation of a mass transfer instability. The above suggests that the disc is very thin during outburst so that there is no significant screening of the flux irradiating the red star. The maximum line flux of H$\beta$ from the red star is 0.4 x 10$^{30}$ erg s$^{-1}$ (the flux density of H$\beta$ is 16.54 mJy from Fig. 1). When compared with an accretion rate of $10^{-9} M_\odot yr^{-1}$ (Z Cha in outburst; Horne & Cook 1985) the luminosity from the boundary layer is roughly 1 x 10$^{34}$ erg s$^{-1}$ with only 1% intercepted by the red star. Assuming an albedo of 0.5 for the red star, the H$\beta$ irradiated line flux is less than the total irradiating flux by about 2 orders of magnitude which shows that there is enough energy for irradiation to work.

Alternatively, the activity on the red star may point to a mass-transfer instability, since it shows not only Balmer emission but also mass transfer through the L$_1$ point, significant enough to be seen through the disc emission. The velocity of the gas stream changes when it finds the accretion disc; from velocity close to ballistic to a velocity closer to keplerian. The anisotropic line distribution below 0.5 $R_{L_1}$ disc is caused by the interaction of the gas stream with the accretion disc. This interaction may represent penetration of the accretion disc by the gas stream resulting to sub-keplerian velocities (by 200 km s$^{-1}$) along the path of the gas stream to a distance close to the white dwarf. The penetration of the disc would then suggest that the gas stream is thicker, enabling it to spill over the accretion disc. Alternatively, the interaction could result in raising part of the outer disc. The arm-like emission in Fig. 5 has velocities 800-1000 km s$^{-1}$ which is consistent with emission from the outer disc. The double-peaked profile in eclipse indicates that the lowest velocities in the outer disc are of the order of 700 km s$^{-1}$.

This is the first case of emission from the gas stream and the red star being observed simultaneously during outburst: its implications on the state of the red star (intrinsic activity, a relation to the outburst mechanism and/or irradiation from a hotter source) are compelling argument for observing similar outbursts in the near future. The origin of the line emission on the red star, which has been seen in systems of all types and outburst states (e.g. IP Peg in outburst by Marsh & Horne 1990 and in quiescence by Harlaftis et al. 1994), has yet to be resolved.

*Acknowledgements.* We thank Keith Horne and Frank Bateson for informing us about the outburst and Frederic Hessman for his useful comments which resulted in considerable improvement of the presentation. ETH was in receipt of a HUMAN CAPITAL AND MOBILITY PROGRAM fellowship CT920216 (1993-1994), funded by the European Union and administered by the Royal Observatories, UK. TRM is supported by a PPARC Advanced Fellowship, UK.

**Fig. 6.** We build a model image (top panel) to simulate the effects of MEM, noise and default image used in the reconstruction technique. The computed data are shown in the middle panel and then the same data with simulated Gaussian noise such that the S/N is similar to that of the H$\beta$ trailed spectra of OY Car. See also text.

6.1. Simulations with a Model gas stream and red star emission

We performed the following simulations with the aim to clarify some aspects of Doppler Tomography. In particular, we wished to show the effect of the technique on the images obtained and finally to give a feeling of how to distinguish between real and artificial structures. We choose a model which comprises of the axisymmetric image of the H$\beta$ data of OY Car and model emission from the red star and gas stream (Fig. 6a). The computed data of this model are built (Fig. 6b) and simulated Gaussian noise is added so that the S/N ratio is similar to the observed data (Fig. 6c). The latter data are then used to make Doppler images. The images are displayed in Fig. 7 where a Gaussian profile of FWHM of 2 and 10 pixels was used for blurring the default in the left and right panels, respectively. Finally, we subtract the axisymmetric parts centred on the white dwarf of OY Car and the results are shown at the bottom panels of Fig. 7.

There, the effects of the Gaussian blurring and noise are most easily assessed. Noise clusters around a few pixels and thereafter interpretation of only the largest or strongest emission formations can be attempted (*e.g.* gas stream and red star). The red star emission (spot image) is reconstructed fully whereas the gas stream reconstruction is blurred from its model 2-pixel width (equal to the velocity resolution). This arises from the combined effect of elongated structure of the gas stream emission (streaklike), the noise and the maximum entropy requirement (the method gives the smoothest solution compatible with the data): compare the top panel of Fig. 6 with the bottom panels of Fig. 7. The Gaussian default further contributes to the blurring the larger the FWHM is. For a FWHM of 10 pixels, the blurring of the reconstructed gas stream is about 4 times the model width. The trajectory of the stream is now identical to the average of the blurred width. A Gaussian is convolved to the default image since it can reconstruct accurately the radial profile of the emission distribution and allow for azimuthal structure to be built. (for a discussion of the Gaussian blurring default versus the uniform default see Marsh & Horne 1988). A FWHM of 10 pixels was used for the Gaussian blurring of the default image since the S/N ratio of the OY Car data was low. This is large enough to allow large scale structure to build up easily but is also smaller than the disc structure. However, we tried Gaussian blurring with a FWHM of 4 pixels and the reconstructed image does not show any essential difference. The Gaussian blurring has eliminated most of the noise content of the image. However, noise-correlated structures have survived but can be used as reference structures to judge between real and artificial structures. The interested reader should also look in the simulations performed by Marsh & Horne (1988) and Harlaftis et al. (1994). Finally, it is interesting to note the shape of the 'S'-wave created by the changing aspect of the model gas stream (Fig. 6).

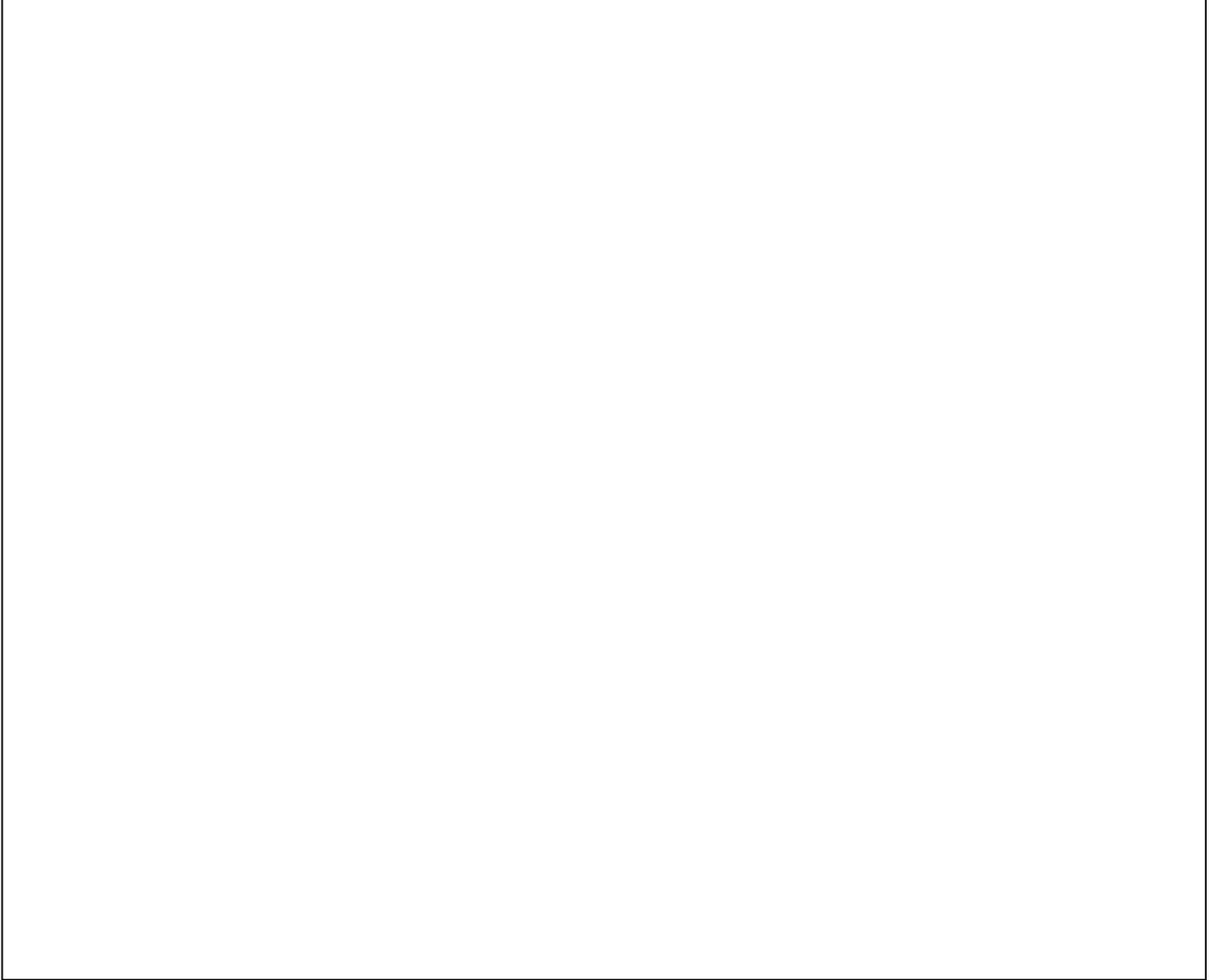

**Fig. 7.** The reconstructed MEM images using the same data (bottom panel of Fig. 6) but different Gaussian blurring for computing the default image (FWHM=2 pixels and 10 pixels from left to right, respectively). The non-axisymmetric parts of the Doppler images are displayed at the bottom panels. The effect of the method is visible, firstly on the gas stream reconstruction and secondly, on the noise content (see also text).

### 6.2. The filtered back-projection Image

We wished to confirm the reliability of the MEM images and against this purpose we reconstructed the H$\beta$ Doppler image using the inverse linear tomography method (back-projection method; see Appendix in Marsh & Horne 1988). The data were first filtered and then back projected. The filter works in the frequency domain by multiplying the Fourier transform of the data with the Fourier transform of the point-spread function $1/V$ of the back-projection method. Noise above a cut-off frequency (less than the Nyquist frequency) is reduced by convolving the filtered data to an appropriate Gaussian (see Horne 1992 for details of the method). The technique has the advantage that the image production is very quick in terms of CPU time and that no treatment is needed for negatives values in the data.

The filtered back-projection image is shown at the top of Fig. 8 with its non-axisymmetric part at the bottom. Comparison with the MEM image of Fig. 5 shows that the main features are consistent, that is the red star with the gas stream emission. This comparison also shows the differences between the two methods which lies in the treatment of noise and that of signal close to the noise the prescribed treatment is the result of its amplification by the back projection method. The filter also has the two-folded effect of cutting the weakest signal and amplifying the strongest. This is evident in the filtered back-projection image which shows that the emission around the $L_1$ point is stronger whereas only half of the arm-like emission is clearly present compared to the MEM image. A check with the unfiltered back-projection image confirms the presence of the arm-like emission down to 0.2 $R_{L_1}$. Thereafter, we conclude that the extent of the MEM arm-like emission is real although around the white dwarf ($< 0.2$ $R_{L_1}$) is comparable to the noise.

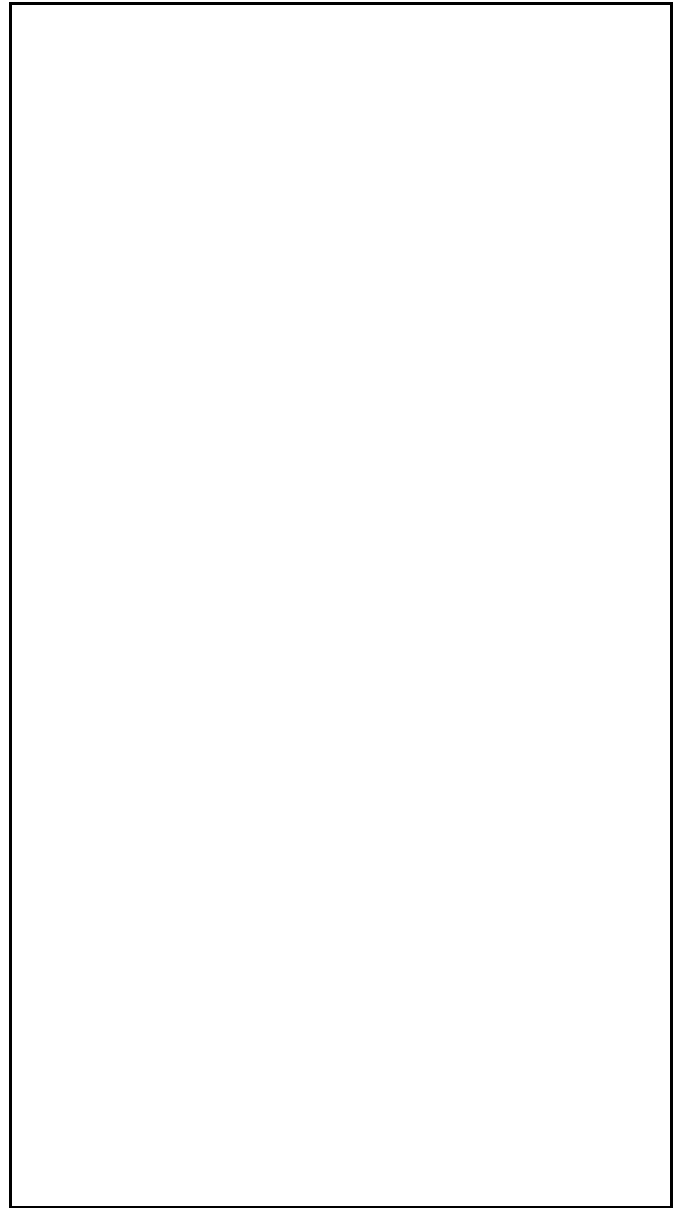

**Fig. 8.** The filtered back-projection image of H$\beta$ is shown at the top. Before the back-projection, the data were convolved with a Gaussian which suppressed high frequency noise above 0.5 times the Nyquist frequency. The bottom panel shows the non-axisymmetric image which can be compared to the MEM image displayed in Fig. 5 (or Fig. 4). The emission distribution is basically the same (see also text).